\documentclass[showkeys, showpacs]{revtex4}
\usepackage{indentfirst}
\begin{document}
\title{\bf Baryonia with open and hidden strange}
\author{S.M. Gerasyuta}
\email{sergei.gerasyuta@mail.ru}
\affiliation{Department of Physics, St. Petersburg State Forest Technical
University, 5 Institutsi Per., St. Petersburg 194021, Russia}
\author{E.E. Matskevich}
\email{Matskevich_EE@pers.spmi.ru}
\affiliation{Department of General and  Technical Physics, St. Petersburg Mining
University, 83 Maliy V.O. Pr., St. Petersburg 199226, Russia}
\begin{abstract}
The relativistic six-quark equations are found in the framework of the
dispersion relation technique. The strange baryonia are constructed
without the mixing of the quarks and antiquarks. The relativistic six-quark
amplitudes of the strange baryonia with the open and hidden strange are
calculated. The poles of these amplitudes determine the masses of strange baryonia.
17 masses of baryonia are predicted.
\end{abstract}
\pacs{11.55.Fv, 11.80.Jy, 12.39.Ki, 12.39.Mk.}
\keywords{Strange baryonia, dispersion relation technique.}
\maketitle
\section{Introduction.}
Theoretical studies of baryon-antibaryon bound states were started by Fermi and Yang
in the study of pions as nucleon-antinucleon pairs \cite{1}.
A Nambu and Jona-Lasinio model \cite{2, 3} was constructed in which the possibility
of obtaining a pion with zero mass as a fermion-antifermion bound state with a doubled
mass of a fermion was considered .

BES Collaboration observed a significant threshold enhancement of $p\bar p$
mass spectrum in the radiative decay $J/\psi\to\gamma\, p\bar p$ \cite{4}.
Recently BES Collaboration reported the results on $X(1835)$
in the $J/\psi\to\gamma\, \eta'\pi^+ \pi^-$ channel \cite{5}. Under the strong
assumption that the $p\bar p$ threshold enhancement and $X(1835)$ are the
same resonance, Zhu and Gao suggested $X(1835)$ could be a $J^{PC}=0^{-+}$
$p\bar p$ baryonium \cite{6}. Enhancement in the baryon-antibaryon channel
near the threshold are expected on the basis of duality arguments \cite{7, 8, 9}
and by comparison with the systematic of resonance formation in meson-meson
and meson-baryon channels \cite{10}. A historical survey of bound states or
resonances coupled to the nucleon-antinucleon channel is given in \cite{11}.
Gluonic states can couple to baryon-antibaryon channels of appropriate spin and parity.

Theoretical work speculated many possibilities for the enhancement
such as the t-channel pion exchange, some kind of threshold kinematical
effects, as new resonance below threshold or $p\bar p$ bound
state \cite{12, 13, 14, 15, 16, 17, 18, 19}.

In Refs. \cite{20, 21, 22, 23, 24} a method has been developed which is
convenient for analysing relativistic three-hadron systems. The physics of
the three-hadron system can be described by means of a pair interaction
between the particles. There are three isobar channels, each of which
consists of a two-particle isobar and the third particle. The presence
of the isobar representation together with the condition of unitarity in
the pair energies and of analyticity leads to a system of integral equations
in a single variable. Their solution makes it possible to describe the
interaction of the produced particles in three-hadron systems.

In Refs. \cite{25, 26, 27} relativistic generalization of the three-body
Faddeev equations was obtained in the form of dispersion relations in the
pair energy of two interacting quarks. The mass spectrum of $S$-wave
baryons including $u$, $d$, $s$ quarks was calculated by a method based on
isolating the leading singularities in the amplitude. We searched for the
approximate solution of integral three-quark equations by taking into
account two-particle and triangle singularities, all the weaker ones being
neglected. If we considered such approximation, which corresponds to
taking into account two-body and triangle singularities, and defined all
the smooth functions of the subenergy variables (as compared with the
singular part of the amplitude) in the middle point of the physical region
of Dalitz-plot, then the problem was reduced to the one of solving a system
of simple algebraic equations.

In Ref. \cite{28} the relativistic six-quark equations are found in
the framework of coupled channel formalism. The dynamical mixing between
the subamplitudes of hexaquark are considered. The six-quark amplitudes
of dibaryons are calculated. The poles of these amplitudes determine the
masses of dibaryons. We calculated the contribution of six-quark
subamplitudes to the hexaquark amplitudes.

In the previous paper \cite{29} the relativistic six-quark equations including $u$, $d$
quarks and antiquarks are found. The nonstrange barionia $B \bar B$ are
constructed without the mixing of the quarks and antiquarks. The
relativistic six-quark amplitudes of the baryonia are calculated.
The poles of these amplitudes determine the masses of baryonia.

In the our paper \cite{30} the charmed barionia are constructed without the mixing
of the quarks and antiquarks. The relativistic six-quark amplitudes of the
heavy baryonia with the open and hidden charm are calculated. The poles of
these amplitudes determine the masses of charmed baryonia.

In the present paper the strange barionia are constructed without the mixing
of the quarks and antiquarks. The relativistic six-quark amplitudes of the
strange baryonia with the open and hidden strange are calculated. The poles of
these amplitudes determine the masses of strange baryonia.

In Sec. II the six-quark amplitudes of baryonia are constructed. The dynamical
mixing between the subamplitudes of baryonia are considered. The relativistic
six-quark equations are obtained in the form of the dispersion relations
over the two-body subenergy. The approximate solutions of these equations
using the method based on the extraction of leading singularities of the
amplitude are obtained. Sec. III is devoted to the calculation results for
the strange baryonia mass spectrum (Tables \ref{tab1}, \ref{tab2}). In conclusion,
the status of the considered model is discussed.
\section{Six-Quark Amplitudes of the Strange Baryonia.}

The relativistic generalization of the three-body Faddeev equations was obtained
in the form of dispersion relations in the pair energy of two interacting quarks.
The pair quarks amplitudes $qq\rightarrow qq$ are calculated in
the framework of the dispersion $N/D$ method with the input four-fermion
interaction with quantum numbers of the gluon \cite{30}.

The construction of the approximate solution is based on
extraction of the leading singularities are close to the region
$s_{ik}\approx 4m^2$. Such a classification of singularities
makes it possible to search for an approximate solution of equations,
taking into account a definite number of leading singularities and
neglecting the weaker ones \cite{28}.

The relativistic six-quark equations in the framework of the dispersion
relation technique are derived. We use only planar diagrams; the other
diagrams due to the rules of $1/N_c$ expansion are neglected.
The current generates a six-quark system. The correct equations for the
amplitude are obtained by taking into account all possible subamplitudes.
Then one should represent a six-particle amplitude as a sum of 15 subamplitudes:

\begin{equation}
\label{1}
A=\sum\limits_{i<j \atop i, j=1}^6 A_{ij}\, . \end{equation}

This defines the division of the diagrams into groups according to the
certain pair interaction of particles. The total amplitude can be
represented graphically as a sum of diagrams. We take into account the
pairwise interaction of all quarks and antiquarks in the baryonia.

We use the results of our relativistic quark model and write down the
pair quark amplitudes in the form:

\begin{equation}
\label{2}
a_n(s_{ik})=\frac{G^2_n(s_{ik})}
{1-B_n(s_{ik})} \, ,\end{equation}

\begin{equation}
\label{3}
B_n(s_{ik})=\int\limits_{(m_i+m_k)^2}^{\frac{(m_i+m_k)^2\Lambda}{4}}
\hskip2mm \frac{ds'_{ik}}{\pi}\frac{\rho_n(s'_{ik})G^2_n(s'_{ik})}
{s'_{ik}-s_{ik}} \, .\end{equation}

\noindent
Here $G_n(s_{ik})$ are the diquark vertex functions (Table \ref{tab3}).
The vertex functions are determined by the contribution of the crossing
channels. The vertex functions satisfy the Fierz relations.

These vertex functions are generated from $g_V$. $B_n(s_{ik})$ and $\rho_n (s_{ik})$
are the Chew-Mandelstam functions with cutoff $\Lambda$ and the phase spaces:

\begin{eqnarray}
\label{4}
\rho_n (s_{ik},J^{PC})&=&\left(\alpha(n,J^{PC}) \frac{s_{ik}}{(m_i+m_k)^2}
+\beta(n,J^{PC})+\delta(n,J^{PC}) \frac{(m_i-m_k)^2}{s_{ik}}\right)
\nonumber\\
&&\nonumber\\
&\times & \frac{\sqrt{(s_{ik}-(m_i+m_k)^2)(s_{ik}-(m_i-m_k)^2)}}
{s_{ik}}\, .
\end{eqnarray}

The coefficients $\alpha(n,J^{PC})$, $\beta(n,J^{PC})$ and
$\delta(n,J^{PC})$ are given in Table \ref{tab3}.

Here $n=1$ corresponds to $q\bar q$-pairs with $J^P=0^-$, $n=2$ corresponds
to the $q\bar q$-pairs with $J^P=1^-$, $n=3$ defines the $qq$-pairs with
$J^P=0^+$, $n=4$ determines $J^P=1^+$ $qq$-pairs.

In the present paper we consider the two types of the six-quark
baryon-antibaryon $B\bar B$ states: the baryon-antibaryons with
one strange quark $qqQ\bar q\bar q\bar q$, where $q=u, d$, $Q=s$,
and the baryon-antibaryons with two strange quarks $qqQ\bar q\bar q\bar Q$,
here $q=u, d, s$, $Q=s$.

The values of quark masses ($m_{u, d}=410\, MeV$, $m_s=557\, MeV$ )
are taken from the previous papers.
We use the parameters of model similar to those in the previous papers:
the gluon coupling constant $g=0.314$ was used in the study of light and charmed
baryonia, the cutoff $\Lambda=11.0$ is usual for the all light six-quarks
states. The cutoffs $\Lambda_{qs}=6.54$ for baryonia with open strange
and $\Lambda_{qs,ss}=9.17$ for baryonia with hidden strange, $q=u, d$
are new parameters.

The results of our calculations are given in the Tables \ref{tab1}, \ref{tab2}.

As the example, we consider the baryonium $\Sigma_s \bar \Delta$
$(uus \, \bar d \bar d \bar d)$ with the spin-parity $J^P=1^-$.
The system of equations for this baryonium is as follows:

\begin{eqnarray}
\label{5}
\alpha_1^{1^{uu}}&=&\lambda+2\, \alpha_1^{0^{us}} I_1(1^{uu}0^{us})
+6\, \alpha_1^{1^{u\bar d}} I_1(1^{uu}1^{u\bar d})
+6\, \alpha_1^{0^{u\bar d}} I_1(1^{uu}0^{u\bar d})\, ,
\\
&&\nonumber\\
\label{6}
\alpha_1^{0^{us}}&=&\lambda+\alpha_1^{1^{uu}} I_1(0^{us}1^{uu})
+\alpha_1^{0^{us}} I_1(0^{us}0^{us})
+3\, \alpha_1^{1^{u\bar d}} I_1(0^{us}1^{u\bar d})
+3\, \alpha_1^{0^{u\bar d}} I_1(0^{us}0^{u\bar d})
\nonumber\\
&&\nonumber\\
&+&3\, \alpha_1^{1^{s\bar d}} I_1(0^{us}1^{s\bar d})
+3\, \alpha_1^{0^{s\bar d}} I_1(0^{us}0^{s\bar d})\, ,
\\
&&\nonumber\\
\label{7}
\alpha_1^{1^{\bar d\bar d}}&=&\lambda+2\, \alpha_1^{1^{\bar d\bar d}} I_1(1^{\bar d\bar d}1^{\bar d\bar d})
+4\, \alpha_1^{1^{u\bar d}} I_1(1^{\bar d\bar d}1^{u\bar d})
+4\, \alpha_1^{0^{u\bar d}} I_1(1^{\bar d\bar d}0^{u\bar d})
+2\, \alpha_1^{1^{s\bar d}} I_1(1^{\bar d\bar d}1^{s\bar d})
\nonumber\\
&&\nonumber\\
&+&2\, \alpha_1^{0^{s\bar d}} I_1(1^{\bar d\bar d}0^{s\bar d})\, ,
\\
&&\nonumber\\
\label{8}
\alpha_1^{1^{u\bar d}}&=&\lambda+\alpha_1^{1^{uu}} I_1(1^{u\bar d}1^{uu})+\alpha_1^{0^{us}} I_1(1^{u\bar d}0^{us})
+2\, \alpha_1^{1^{\bar d\bar d}} I_1(1^{u\bar d}1^{\bar d\bar d})
+3\, \alpha_1^{1^{u\bar d}} I_1(1^{u\bar d}1^{u\bar d})
\nonumber\\
&&\nonumber\\
&+&3\, \alpha_1^{0^{u\bar d}} I_1(1^{u\bar d}0^{u\bar d})
+\alpha_1^{1^{s\bar d}} I_1(1^{u\bar d}1^{s\bar d})+\alpha_1^{0^{s\bar d}} I_1(1^{u\bar d}0^{s\bar d})
+2\, \alpha_2^{1^{uu}1^{\bar d\bar d}} I_2(1^{u\bar d}1^{uu}1^{\bar d\bar d})
\nonumber\\
&&\nonumber\\
&+&2\, \alpha_2^{0^{us}1^{\bar d\bar d}} I_2(1^{u\bar d}0^{us}1^{\bar d\bar d})\, ,
\\
&&\nonumber\\
\label{9}
\alpha_1^{0^{u\bar d}}&=&\lambda+\alpha_1^{1^{uu}} I_1(0^{u\bar d}1^{uu})+\alpha_1^{0^{us}} I_1(0^{u\bar d}0^{us})
+2\, \alpha_1^{1^{\bar d\bar d}} I_1(0^{u\bar d}1^{\bar d\bar d})
+3\, \alpha_1^{1^{u\bar d}} I_1(0^{u\bar d}1^{u\bar d})
\nonumber\\
&&\nonumber\\
&+&3\, \alpha_1^{0^{u\bar d}} I_1(0^{u\bar d}0^{u\bar d})
+\alpha_1^{1^{s\bar d}} I_1(0^{u\bar d}1^{s\bar d})+\alpha_1^{0^{s\bar d}} I_1(0^{u\bar d}0^{s\bar d})
+2\, \alpha_2^{1^{uu}1^{\bar d\bar d}} I_2(0^{u\bar d}1^{uu}1^{\bar d\bar d})
\nonumber\\
&&\nonumber\\
&+&2\, \alpha_2^{0^{us}1^{\bar d\bar d}} I_2(0^{u\bar d}0^{us}1^{\bar d\bar d})\, ,
\\
&&\nonumber\\
\label{10}
\alpha_1^{1^{s\bar d}}&=&\lambda+2\, \alpha_1^{1^{\bar d\bar d}} I_1(1^{s\bar d}1^{\bar d\bar d})
+2\, \alpha_1^{1^{u\bar d}} I_1(1^{s\bar d}1^{u\bar d})
+2\, \alpha_1^{0^{u\bar d}} I_1(1^{s\bar d}0^{u\bar d})
+2\, \alpha_1^{1^{s\bar d}} I_1(1^{s\bar d}1^{s\bar d})
\nonumber\\
&&\nonumber\\
&+&2\, \alpha_1^{0^{s\bar d}} I_1(1^{s\bar d}0^{s\bar d})
+4\, \alpha_2^{0^{us}1^{\bar d\bar d}} I_2(1^{s\bar d}0^{us}1^{\bar d\bar d})\, ,
\\
&&\nonumber\\
\label{11}
\alpha_1^{0^{s\bar d}}&=&\lambda+2\, \alpha_1^{1^{\bar d\bar d}} I_1(0^{s\bar d}1^{\bar d\bar d})
+2\, \alpha_1^{1^{u\bar d}} I_1(0^{s\bar d}1^{u\bar d})
+2\, \alpha_1^{0^{u\bar d}} I_1(0^{s\bar d}0^{u\bar d})
+2\, \alpha_1^{1^{s\bar d}} I_1(0^{s\bar d}1^{s\bar d})
\nonumber\\
&&\nonumber\\
&+&2\, \alpha_1^{0^{s\bar d}} I_1(0^{s\bar d}0^{s\bar d})
+4\, \alpha_2^{0^{us}1^{\bar d\bar d}} I_2(0^{s\bar d}0^{us}1^{\bar d\bar d})\, ,
\\
&&\nonumber\\
\label{12}
\alpha_2^{1^{uu}1^{\bar d\bar d}}&=&\lambda+2\, \alpha_1^{0^{us}} I_4(1^{uu}1^{\bar d\bar d}0^{us})
+2\, \alpha_1^{1^{\bar d\bar d}} I_4(1^{\bar d\bar d}1^{uu}1^{\bar d\bar d})
+4\, \alpha_1^{1^{u\bar d}} I_3(1^{uu}1^{\bar d\bar d}1^{u\bar d})
\nonumber\\
&&\nonumber\\
&+&4\, \alpha_1^{0^{u\bar d}} I_3(1^{uu}1^{\bar d\bar d}0^{u\bar d})
+4\, \alpha_2^{0^{us}1^{\bar d\bar d}} I_6(1^{uu}1^{\bar d\bar d}0^{us}1^{\bar d\bar d})\, ,
\\
&&\nonumber\\
\label{13}
\alpha_2^{0^{us}1^{\bar d\bar d}}&=&\lambda+\alpha_1^{1^{uu}} I_4(0^{us}1^{\bar d\bar d}1^{uu})
+\alpha_1^{0^{us}} I_4(0^{us}1^{\bar d\bar d}0^{us})
+2\, \alpha_1^{1^{\bar d\bar d}} I_4(1^{\bar d\bar d}0^{us}1^{\bar d\bar d})
\nonumber\\
&&\nonumber\\
&+&2\, \alpha_1^{1^{u\bar d}} I_3(0^{us}1^{\bar d\bar d}1^{u\bar d})
+2\, \alpha_1^{0^{u\bar d}} I_3(0^{us}1^{\bar d\bar d}0^{u\bar d})
+2\, \alpha_1^{1^{s\bar d}} I_3(0^{us}1^{\bar d\bar d}1^{s\bar d})
\nonumber\\
&&\nonumber\\
&+&2\, \alpha_1^{0^{s\bar d}} I_3(0^{us}1^{\bar d\bar d}0^{s\bar d})
+2\, \alpha_2^{1^{uu}1^{\bar d\bar d}} I_6(0^{us}1^{\bar d\bar d}1^{uu}1^{\bar d\bar d})
+2\, \alpha_2^{0^{us}1^{\bar d\bar d}} I_6(0^{us}1^{\bar d\bar d}0^{us}1^{\bar d\bar d})\, .
\end{eqnarray}

For simplicity, the amplitudes $\alpha_3$ are neglected.

In Fig. 1 the graphical representation of equation (\ref{12}) is given.

\section{Calculation results.}

In Fig. 1 the first and the second coefficients are equal to 2, that is, the
permutation of particles 1 and 2; the third and the fourth coefficients are
equal to 4, that is, the number $4=2$ (the permutation of particles 1 and 2)
$\times 2$ (the permutation of particles 3 and 4); the fifth coefficient is
equal to 4, that is, the number $4=2$ (the permutation of particles 1 and 2)
$\times 2$ (the permutation of particles 3 and 4).

The similar approach allows us to take into account the coefficients
in all the equations.

The poles of the reduced amplitudes $\alpha_l$ correspond to the bound states
and determine the masses of the strange baryonia.

We consider baryonia with the content $qqQ\bar q\bar q\bar q$ and the
spin-parities $J^P=0^-$, $1^-$, $2^-$. The isospin projections are
equal to $\frac{1}{2}$, $\frac{3}{2}$, $\frac{5}{2}$ (Table \ref{tab1}).

The degeneration of baryonium masses with the different spin-parities
$J^P=0^-$, $1^-$ was obtained. We cannot also calculate the bound states
of baryonia with $J^P=3^-$.

The baryonium state $\Sigma_s \bar \Delta$ $(uus \, \bar d \bar d \bar d)$
for the spin-parities $J^P=0^-$, $1^-$, $2^-$ is calculated with the nine
subamplitudes: seven $\alpha_1$ (similar to $\alpha_1^{1^{uu}}$) and
two $\alpha_2^{1^{uu}1^{\bar d \bar d}}$, $\alpha_2^{0^{us}1^{\bar d \bar d}}$.

The baryonium $\Sigma_s \bar \Delta$ $(uus \, \bar u \bar d \bar d)$
consists of 16 subamplitudes with the spin-parities $J^P=0^-$, $1^-$;
12 $\alpha_1$ and 4 $\alpha_2$: $\alpha_2^{1^{uu}1^{\bar d \bar d}}$,
$\alpha_2^{1^{uu}0^{\bar u \bar d}}$, $\alpha_2^{0^{us}1^{\bar d \bar d}}$,
$\alpha_2^{0^{us}0^{\bar u \bar d}}$. For the case of the spin-parity
$J^P=2^-$ the subamplitude $\alpha_2^{0^{us}0^{\bar u \bar d}}$ is
absent. The states with spin-parities $J^P=0^-$, $1^-$, $2^-$
$(uds \, \bar u \bar u \bar u)$ are constructed with 13 subamplitudes:
10 $\alpha_1$ and 3 $\alpha_2$: $\alpha_2^{0^{ud}1^{\bar u \bar u}}$,
$\alpha_2^{0^{us}1^{\bar u \bar u}}$, $\alpha_2^{0^{ds}1^{\bar u \bar u}}$.
The baryonium $uds \, \bar u \bar u \bar d$ for the spin-parities
$J^P=0^-$, $1^-$ takes into account 23 subamplitudes: 17 $\alpha_1$ and
6 $\alpha_2$; $\alpha_2^{0^{ud}0^{\bar u \bar d}}$,
$\alpha_2^{0^{us}0^{\bar u \bar d}}$, $\alpha_2^{0^{ds}0^{\bar u \bar d}}$,
$\alpha_2^{0^{ud}1^{\bar u \bar u}}$, $\alpha_2^{0^{us}1^{\bar u \bar u}}$,
$\alpha_2^{0^{ds}1^{\bar u \bar u}}$. For the case $J^P=2^-$ the
subamplitudes $\alpha_2^{0^{ud}0^{\bar u \bar d}}$,
$\alpha_2^{0^{us}0^{\bar u \bar d}}$, $\alpha_2^{0^{ds}0^{\bar u \bar d}}$
are absent.

We predict the mass of lowest open strange baryonium with the isospin projection
$I_3=\frac{1}{2}$ and the spin-parity $J^P=1^-$ ($M=2085\, MeV$).

We also predict the masses of strange baryonium with the isospin projection
$I_3=\frac{1}{2},\frac{3}{2}$ and the spin-parity $J^P=0^-$ $M=2100\, MeV$ $\Gamma=33\, MeV$,
$J^P=1^-$ $M=2100\, MeV$ $\Gamma=33\, MeV$,
and $I_3=\frac{1}{2}$ $J^P=0^-$ $M=2110\, MeV$ $\Gamma=23\, MeV$,
$J^P=1^-$ $M=2110\, MeV$ $\Gamma=23\, MeV$.
These states have a small width with respect to their masses.

The baryonium $N\bar \Sigma_s$ $(uud \, \bar u \bar u \bar s)$
consists of 16 subamplitudes with the spin-parities $J^P=0^-$, $1^-$;
12 $\alpha_1$ and 4 $\alpha_2$: $\alpha_2^{1^{uu}1^{\bar u \bar u}}$,
$\alpha_2^{1^{uu}0^{\bar u \bar s}}$, $\alpha_2^{0^{ud}1^{\bar u \bar u}}$,
$\alpha_2^{0^{ud}0^{\bar u \bar s}}$.
The baryonium $N\bar \Sigma_s$ $(uud \, \bar u \bar d \bar s)$
consists of 23 subamplitudes with the spin-parities $J^P=0^-$, $1^-$;
17 $\alpha_1$ and 6 $\alpha_2$: $\alpha_2^{1^{uu}0^{\bar u \bar d}}$,
$\alpha_2^{1^{uu}0^{\bar u \bar s}}$, $\alpha_2^{1^{uu}0^{\bar d \bar s}}$,
$\alpha_2^{0^{ud}0^{\bar u \bar d}}$,
$\alpha_2^{0^{ud}0^{\bar u \bar s}}$, $\alpha_2^{0^{ud}0^{\bar d \bar s}}$.

But we can see that the width small as compared the $B\bar B$ state (1835).
The charm baryonia $I_3=0;1$ $J^P=0^-,1^-$ $M=4893\, MeV$ $\Gamma=45\, MeV$

The baryonia with the content $qqQ\bar q\bar q\bar Q$ and the
spin-parities $J^P=0^-$, $1^-$, $2^-$ are considered. The isospin projections are
equal to 0, 1, 2 (Table \ref{tab2}).

The baryonium $\Lambda_s \bar \Lambda_s$ ($uds\,\, \bar u\bar d\bar s$) is
calculated with the 33 subamplitudes (equations), 24 $\alpha_1$ (for instance,
$\alpha_1^{0^{ud}}$) and 9 $\alpha_2$: $\alpha_2^{0^{ud}0^{\bar u\bar d}}$,
$\alpha_2^{0^{ud}0^{\bar u\bar s}}$, $\alpha_2^{0^{ud}0^{\bar d\bar s}}$,
$\alpha_2^{0^{us}0^{\bar u\bar d}}$, $\alpha_2^{0^{us}0^{\bar u\bar s}}$,
$\alpha_2^{0^{us}0^{\bar d\bar s}}$, $\alpha_2^{0^{ds}0^{\bar u\bar d}}$,
$\alpha_2^{0^{ds}0^{\bar u\bar s}}$, $\alpha_2^{0^{ds}0^{\bar d\bar s}}$.
The isospin projection is equal to $I_3=0$ and the spin-parities $J^P=0^-$, $1^-$.
We predict the degeneracy of strange baryonia (Table \ref{tab2}) with the
mass $M=2200\, MeV$.
These states also have a small width with respect to their masses, $\Gamma=32\, MeV$.

The model in question the baryonia $\Sigma_s \bar \Sigma_s$ ($uus\,\, \bar u\bar u\bar s$)
is described with 16 subamplitudes, 12 $\alpha_1$ and 4 $\alpha_2$:
$\alpha_2^{1^{uu}1^{\bar u\bar u}}$, $\alpha_2^{1^{uu}0^{\bar u\bar s}}$,
$\alpha_2^{0^{us}1^{\bar u\bar u}}$, $\alpha_2^{0^{us}0^{\bar u\bar s}}$
for the spin-parities $J^P=0^-$, $1^-$. (Baryonium mass $M=2180\, MeV$).
In the case $J^P=2^-$ we considered 15 subamplitudes:
12 $\alpha_1$ and 3 $\alpha_2$: $\alpha_2^{1^{uu}1^{\bar u\bar u}}$,
$\alpha_2^{1^{uu}0^{\bar u\bar s}}$, $\alpha_2^{0^{us}1^{\bar u\bar u}}$.
(Baryonium mass $M=2189\, MeV$).

The baryonium $\Sigma_s \bar \Sigma_s$ $uds\,\, \bar u\bar u\bar s$ in the case of
spin-parities $J^P=0^-$, $1^-$ is calculated with the 23 subamplitudes (equations),
17 $\alpha_1$ and 6 $\alpha_2$: $\alpha_2^{0^{ud}1^{\bar u\bar u}}$,
$\alpha_2^{0^{us}1^{\bar u\bar u}}$, $\alpha_2^{0^{ds}1^{\bar u\bar u}}$,
$\alpha_2^{0^{ud}0^{\bar u\bar s}}$, $\alpha_2^{0^{us}0^{\bar u\bar s}}$,
$\alpha_2^{0^{ds}0^{\bar u\bar s}}$. (Baryonium mass $M=2190\, MeV$).
For the spin-parity $J^P=2^-$ we used the 20 subamplitudes: 17 $\alpha_1$ and
3 $\alpha_2$: $\alpha_2^{0^{ud}1^{\bar u\bar u}}$, $\alpha_2^{0^{us}1^{\bar u\bar u}}$,
$\alpha_2^{0^{ds}1^{\bar u\bar u}}$. (Baryonium mass $M=2205\, MeV$).

We used the functions
$I_1$, $I_2$, $I_3$, $I_4$, $I_5$, $I_6$:

\begin{eqnarray}
\label{14}
I_1(ij)&=&\frac{B_j(s_0^{13})}{B_i(s_0^{12})}
\int\limits_{(m_1+m_2)^2}^{\frac{(m_1+m_2)^2\Lambda_i}{4}}
\frac{ds'_{12}}{\pi}\frac{G_i^2(s_0^{12})\rho_i(s'_{12})}
{s'_{12}-s_0^{12}} \int\limits_{-1}^{+1} \frac{dz_1(1)}{2}
\frac{1}{1-B_j (s'_{13})}\, , \\
&&\nonumber\\
\label{15}
I_2(ijk)&=&\frac{B_j(s_0^{13}) B_k(s_0^{24})}{B_i(s_0^{12})}
\int\limits_{(m_1+m_2)^2}^{\frac{(m_1+m_2)^2\Lambda_i}{4}}
\frac{ds'_{12}}{\pi}\frac{G_i^2(s_0^{12})\rho_i(s'_{12})}
{s'_{12}-s_0^{12}}
\frac{1}{2\pi}\int\limits_{-1}^{+1}\frac{dz_1(2)}{2}
\int\limits_{-1}^{+1} \frac{dz_2(2)}{2}\nonumber\\
&&\nonumber\\
&\times&
\int\limits_{z_3(2)^-}^{z_3(2)^+} dz_3(2)
\frac{1}{\sqrt{1-z_1^2(2)-z_2^2(2)-z_3^2(2)+2z_1(2) z_2(2) z_3(2)}}
\nonumber\\
&&\nonumber\\
&\times& \frac{1}{1-B_j (s'_{13})} \frac{1}{1-B_k (s'_{24})}
 \, , \\
&&\nonumber\\
\label{16}
I_3(ijk)&=&\frac{B_k(s_0^{23})}{B_i(s_0^{12}) B_j(s_0^{34})}
\int\limits_{(m_1+m_2)^2}^{\frac{(m_1+m_2)^2\Lambda_i}{4}}
\frac{ds'_{12}}{\pi}\frac{G_i^2(s_0^{12})\rho_i(s'_{12})}
{s'_{12}-s_0^{12}}\nonumber\\
&&\nonumber\\
&\times&\int\limits_{(m_3+m_4)^2}^{\frac{(m_3+m_4)^2\Lambda_j}{4}}
\frac{ds'_{34}}{\pi}\frac{G_j^2(s_0^{34})\rho_j(s'_{34})}
{s'_{34}-s_0^{34}}
\int\limits_{-1}^{+1} \frac{dz_1(3)}{2} \int\limits_{-1}^{+1}
\frac{dz_2(3)}{2} \frac{1}{1-B_k (s'_{23})} \, , \\
&&\nonumber\\
\label{17}
I_4(ijk)&=&I_1(ik) \, , \\
&&\nonumber\\
\label{18}
I_5(ijkl)&=&I_2(ikl) \, , \\
&&\nonumber\\
\label{19}
I_6(ijkl)&=&I_1(ik) \cdot I_1(jl)
 \, .
\end{eqnarray}

\noindent
Here $i$, $j$, $k$, $l$, $m$ correspond to the diquarks with the
spin-parity $J^P=0^+, 1^+$.

\section{Conclusions.}

We calculated the masses of strange baryonia (Tables \ref{tab1}, \ref{tab2}). We predicted 17 masses and
the degeneration of some states.

We obtain that these states have the widths depended of heavy quarks.
We have obtained that some strange and charmed states possess small width.

We predict the mass of lowest open strange baryonium with the isospin projection
$I_3=\frac{1}{2}$ and the spin-parity $J^P=1^-$ ($M=2085\, MeV$).

We also predict the masses of strange baryonium with the isospin projection
$I_3=\frac{1}{2},\frac{3}{2}$ and the spin-parity $J^P=0^-$ $M=2100\, MeV$ $\Gamma=33\, MeV$,
$J^P=1^-$ $M=2100\, MeV$ $\Gamma=33\, MeV$,
and $I_3=\frac{1}{2}$ $J^P=0^-$ $M=2110\, MeV$ $\Gamma=23\, MeV$,
$J^P=1^-$ $M=2110\, MeV$ $\Gamma=23\, MeV$.
These states have a small width with respect to their masses.

\begin{acknowledgments}
The authors would like to thank T. Barnes for useful discussions.
\end{acknowledgments}

\newpage

\vskip80pt

\begin{picture}(600,100)
\put(0,45){\line(1,0){18}}
\put(0,47){\line(1,0){17.5}}
\put(0,49){\line(1,0){17}}
\put(0,51){\line(1,0){17}}
\put(0,53){\line(1,0){17.5}}
\put(0,55){\line(1,0){18}}
\put(30,50){\circle{25}}
\put(19,46){\line(1,1){15}}
\put(22,41){\line(1,1){17}}
\put(27.5,38.5){\line(1,1){14}}
\put(31,63){\vector(1,1){20}}
\put(31,38){\vector(1,-1){20}}
\put(47.5,60){\circle{16}}
\put(47.5,40){\circle{16}}
\put(55,64){\vector(3,2){18}}
\put(55,36){\vector(3,-2){18}}
\put(55,64){\vector(3,-2){18}}
\put(55,36){\vector(3,2){18}}
\put(78,75){$u$}
\put(63,75){\small 1}
\put(78,53){$u$}
\put(70,58){\small 2}
\put(78,41){$\bar d$}
\put(70,36){\small 3}
\put(78,18){$\bar d$}
\put(63,18){\small 4}
\put(54,80){$s$}
\put(41,80){\small 5}
\put(54,13){$\bar d$}
\put(41,13){\small 6}
\put(40.5,56){\footnotesize $1^{uu}$}
\put(40.5,36){\footnotesize $1^{\bar d \bar d}$}
\put(90,48){$=$}
\put(25,-10){$\alpha_2^{1^{uu}1^{\bar d\bar d}}$}
\put(110,45){\line(1,0){19}}
\put(110,47){\line(1,0){21}}
\put(110,49){\line(1,0){23}}
\put(110,51){\line(1,0){23}}
\put(110,53){\line(1,0){21}}
\put(110,55){\line(1,0){19}}
\put(140,60){\circle{16}}
\put(140,40){\circle{16}}
\put(147.5,64){\vector(3,2){18}}
\put(147.5,36){\vector(3,-2){18}}
\put(147.5,64){\vector(3,-2){18}}
\put(147.5,36){\vector(3,2){18}}
\put(128,55){\vector(1,3){11}}
\put(128,45){\vector(1,-3){11}}
\put(170,75){$u$}
\put(155,75){\small 1}
\put(170,53){$u$}
\put(159,59){\small 2}
\put(170,41){$\bar d$}
\put(159,35){\small 3}
\put(170,18){$\bar d$}
\put(155,18){\small 4}
\put(143,86){$s$}
\put(128,84){\small 5}
\put(143,08){$\bar d$}
\put(128,10){\small 6}
\put(133,57){\footnotesize $1^{uu}$}
\put(133,37){\footnotesize $1^{\bar d \bar d}$}
\put(183,48){$+$}
\put(130,-10){$\lambda$}
\put(199,48){2}
\put(212,45){\line(1,0){18}}
\put(212,47){\line(1,0){17.5}}
\put(212,49){\line(1,0){17}}
\put(212,51){\line(1,0){17}}
\put(212,53){\line(1,0){17.5}}
\put(212,55){\line(1,0){18}}
\put(242,50){\circle{25}}
\put(231,46){\line(1,1){15}}
\put(234,41){\line(1,1){17}}
\put(239.5,38.5){\line(1,1){14}}
\put(258,62){\circle{16}}
\put(263.5,68.5){\vector(1,1){15}}
\put(263.5,68.5){\vector(1,-1){18}}
\put(254,50){\vector(1,0){28}}
\put(290,50){\circle{16}}
\put(298,50){\vector(3,1){22}}
\put(298,50){\vector(3,-1){22}}
\put(274,61){$u$}
\put(269,65){\small 1}
\put(271,41){$u$}
\put(269,52){\small 2}
\put(314,62){$u$}
\put(304,59){\small 1}
\put(314,30){$u$}
\put(304,34){\small 2}
\put(280,85){$s$}
\put(268,83){\small 5}
\put(258,37){\circle{16}}
\put(265,32){\vector(3,-1){20}}
\put(265,32){\vector(2,-3){12}}
\put(243,38){\vector(1,-3){8}}
\put(288,22){$\bar d$}
\put(278,32){\small 3}
\put(279,7){$\bar d$}
\put(269,7){\small 4}
\put(253,7){$\bar d$}
\put(241,9){\small 6}
\put(251,59){\footnotesize $0^{us}$}
\put(283,47){\footnotesize $1^{uu}$}
\put(251,34){\footnotesize $1^{\bar d \bar d}$}
\put(339,48){$+$}
\put(230,-10){$2\, \alpha_1^{0^{us}}\, I_4(1^{uu}1^{\bar d \bar d}0^{us})$}
\put(355,48){2}
\put(368,45){\line(1,0){18}}
\put(368,47){\line(1,0){17.5}}
\put(368,49){\line(1,0){17}}
\put(368,51){\line(1,0){17}}
\put(368,53){\line(1,0){17.5}}
\put(368,55){\line(1,0){18}}
\put(398,50){\circle{25}}
\put(387,46){\line(1,1){15}}
\put(390,41){\line(1,1){17}}
\put(395.5,38.5){\line(1,1){14}}
\put(414,62){\circle{16}}
\put(419.5,68.5){\vector(1,1){15}}
\put(419.5,68.5){\vector(1,-1){18}}
\put(410,50){\vector(1,0){28}}
\put(446,50){\circle{16}}
\put(454,50){\vector(3,1){22}}
\put(454,50){\vector(3,-1){22}}
\put(430,61){$\bar d$}
\put(425,65){\small 1}
\put(427,41){$\bar d$}
\put(425,52){\small 2}
\put(470,62){$\bar d$}
\put(460,59){\small 1}
\put(470,30){$\bar d$}
\put(460,34){\small 2}
\put(436,85){$\bar d$}
\put(424,83){\small 5}
\put(414,37){\circle{16}}
\put(421,32){\vector(3,-1){20}}
\put(421,32){\vector(2,-3){12}}
\put(399,38){\vector(1,-3){8}}
\put(444,22){$u$}
\put(434,32){\small 3}
\put(435,7){$u$}
\put(425,7){\small 4}
\put(409,7){$s$}
\put(397,9){\small 6}
\put(407,59){\footnotesize $1^{\bar d \bar d}$}
\put(439,47){\footnotesize $1^{\bar d \bar d}$}
\put(407,34){\footnotesize $1^{uu}$}
\put(385,-10){$2\, \alpha_1^{1^{\bar d \bar d}}\, I_4(1^{\bar d \bar d}1^{uu}1^{\bar d \bar d})$}
\end{picture}

\vskip80pt
\begin{picture}(600,60)
\put(90,48){$+$}
\put(107,48){4}
\put(125,45){\line(1,0){18}}
\put(125,47){\line(1,0){17.5}}
\put(125,49){\line(1,0){17}}
\put(125,51){\line(1,0){17}}
\put(125,53){\line(1,0){17.5}}
\put(125,55){\line(1,0){18}}
\put(155,50){\circle{25}}
\put(144,46){\line(1,1){15}}
\put(147,41){\line(1,1){17}}
\put(152.5,38.5){\line(1,1){14}}
\put(156,63){\vector(1,1){20}}
\put(156,38){\vector(1,-1){20}}
\put(179,80){$s$}
\put(165,80){\small 5}
\put(179,13){$\bar d$}
\put(165,13){\small 6}
\put(175,50){\circle{16}}
\put(183,50){\vector(3,2){17}}
\put(183,50){\vector(3,-2){17}}
\put(160,61){\vector(1,0){40}}
\put(160,39){\vector(1,0){40}}
\put(208,60){\circle{16}}
\put(208,40){\circle{16}}
\put(216,61){\vector(3,1){20}}
\put(216,61){\vector(3,-1){20}}
\put(216,39){\vector(3,1){20}}
\put(216,39){\vector(3,-1){20}}
\put(241,70){$u$}
\put(226,70){\small 1}
\put(241,53){$u$}
\put(223,51){\small 2}
\put(241,41){$\bar d$}
\put(223,44){\small 3}
\put(241,24){$\bar d$}
\put(226,24){\small 4}
\put(183,66){$u$}
\put(173,65){\small 1}
\put(188,49){\small $u$}
\put(183,54){\small 2}
\put(193,45){\small $\bar d$}
\put(183,40){\small 3}
\put(183,27){$\bar d$}
\put(173,28){\small 4}
\put(168,47){\footnotesize $1^{u \bar d}$}
\put(201,57){\footnotesize $1^{uu}$}
\put(201,37){\footnotesize $1^{\bar d \bar d}$}
\put(140,-10){$4\, \alpha_1^{1^{u \bar d}}\, I_3(1^{uu}1^{\bar d \bar d}1^{u \bar d})$}
\put(253,48){$+$}
\put(270,48){4}
\put(285,45){\line(1,0){18}}
\put(285,47){\line(1,0){17.5}}
\put(285,49){\line(1,0){17}}
\put(285,51){\line(1,0){17}}
\put(285,53){\line(1,0){17.5}}
\put(285,55){\line(1,0){18}}
\put(315,50){\circle{25}}
\put(304,46){\line(1,1){15}}
\put(307,41){\line(1,1){17}}
\put(312.5,38.5){\line(1,1){14}}
\put(316,63){\vector(1,1){20}}
\put(316,38){\vector(1,-1){20}}
\put(339,80){$s$}
\put(325,80){\small 5}
\put(339,13){$\bar d$}
\put(325,13){\small 6}
\put(335,50){\circle{16}}
\put(343,50){\vector(3,2){17}}
\put(343,50){\vector(3,-2){17}}
\put(320,61){\vector(1,0){40}}
\put(320,39){\vector(1,0){40}}
\put(368,60){\circle{16}}
\put(368,40){\circle{16}}
\put(376,61){\vector(3,1){20}}
\put(376,61){\vector(3,-1){20}}
\put(376,39){\vector(3,1){20}}
\put(376,39){\vector(3,-1){20}}
\put(401,70){$u$}
\put(386,70){\small 1}
\put(401,53){$u$}
\put(383,51){\small 2}
\put(401,41){$\bar d$}
\put(383,44){\small 3}
\put(401,24){$\bar d$}
\put(386,24){\small 4}
\put(343,66){$u$}
\put(333,65){\small 1}
\put(348,49){\small $u$}
\put(343,54){\small 2}
\put(353,45){\small $\bar d$}
\put(343,40){\small 3}
\put(343,27){$\bar d$}
\put(333,28){\small 4}
\put(328,47){\footnotesize $0^{u \bar d}$}
\put(361,57){\footnotesize $1^{uu}$}
\put(361,37){\footnotesize $1^{\bar d \bar d}$}
\put(300,-10){$4\, \alpha_1^{0^{u \bar d}}\, I_3(1^{uu}1^{\bar d \bar d}0^{u \bar d})$}
\end{picture}

\vskip80pt
\begin{picture}(600,60)
\put(90,48){$+$}
\put(107,48){4}
\put(125,45){\line(1,0){18}}
\put(125,47){\line(1,0){17.5}}
\put(125,49){\line(1,0){17}}
\put(125,51){\line(1,0){17}}
\put(125,53){\line(1,0){17.5}}
\put(125,55){\line(1,0){18}}
\put(155,50){\circle{25}}
\put(144,46){\line(1,1){15}}
\put(147,41){\line(1,1){17}}
\put(152.5,38.5){\line(1,1){14}}
\put(164,68){\circle{16}}
\put(164,32){\circle{16}}
\put(167,53){\vector(3,1){28}}
\put(167,47){\vector(3,-1){28}}
\put(172,70){\vector(3,2){21}}
\put(172,70){\vector(3,-1){23}}
\put(172,30){\vector(3,1){23}}
\put(172,30){\vector(3,-2){21}}
\put(203,60){\circle{16}}
\put(203,40){\circle{16}}
\put(212,60){\vector(3,2){21}}
\put(212,60){\vector(3,-1){23}}
\put(212,40){\vector(3,1){23}}
\put(212,40){\vector(3,-2){21}}
\put(239,71){$u$}
\put(224,74){\small 1}
\put(240,53){$u$}
\put(230,57){\small 2}
\put(240,41){$\bar d$}
\put(230,38){\small 3}
\put(239,19){$\bar d$}
\put(224,19){\small 4}
\put(183,86){$s$}
\put(173,82){\small 5}
\put(183,6){$\bar d$}
\put(173,11){\small 6}
\put(188,68){$u$}
\put(182,68){\small 1}
\put(182,51){$u$}
\put(176,59){\small 2}
\put(188,42){$\bar d$}
\put(178,45){\small 3}
\put(188,23){$\bar d$}
\put(181,26){\small 4}
\put(157,65){\footnotesize $0^{us}$}
\put(157,29){\footnotesize $1^{\bar d \bar d}$}
\put(196,57){\footnotesize $1^{uu}$}
\put(196,37){\footnotesize $1^{\bar d \bar d}$}
\put(130,-10){$4\, \alpha_2^{0^{us}1^{\bar d \bar d}}\, I_6(1^{uu}1^{\bar d \bar d}0^{us}1^{\bar d \bar d})$}
\end{picture}

\vskip60pt

Fig. 1. The graphical equations of the reduced amplitude $\alpha_2^{1^{uu}1^{\bar d\bar d}}$
with the projection of isospin $I_3=\frac{1}{2}$, $\frac{3}{2}$ and the spin-parity $J^P=1^-$
$\Sigma_s\bar\Delta (uus\, \bar d\bar d\bar d)$.

\begin{table}
\caption{$qqQ\bar q\bar q\bar q$, $q=u, d$, $Q=s$. Parameters of model: cutoff
$\Lambda=11.0$, $\Lambda_{qs}=6.54$, gluon coupling constant $g=0.314$.
Quark masses $m_q=410\, MeV$, $m_c=557\, MeV$.}
\label{tab1}
\begin{tabular}{|c|c|c|c|c|c|}
\hline
Quark content & $I_3$ & $J$ & Baryonium & Mass (MeV) & Binding energy (MeV) \\
\hline
$uus\,\, \bar u\bar u\bar u$, $dds\,\, \bar d\bar d\bar d$, & $\frac{1}{2}$; $\frac{5}{2}$ & 0 &
$\Sigma^*_s \bar \Delta$, $\Delta \bar \Sigma^*_s$ & 2092 & 525 \\
$uuu\,\, \bar u\bar u\bar s$, $ddd\,\, \bar d\bar d\bar s$; &  & 1 &
$\Sigma_s \bar \Delta$, $\Delta \bar \Sigma_s$ & 2085 & 340 \\
$uus\,\, \bar d\bar d\bar d$, $dds\,\, \bar u\bar u\bar u$, &  &  &
$\Sigma^*_s \bar \Delta$, $\Delta \bar \Sigma^*_s$ & 2085 & 532 \\
$ddd\,\, \bar u\bar u\bar s$, $uuu\,\, \bar d\bar d\bar s$ & & 2 &
$\Sigma_s \bar \Delta$, $\Delta \bar \Sigma_s$ & 2091 & 334 \\
 & & &
$\Sigma^*_s \bar \Delta$, $\Delta \bar \Sigma^*_s$ & 2091 & 526 \\
\hline
$uus\,\, \bar u\bar u\bar d$, $dds\,\, \bar u\bar d\bar d$, & $\frac{1}{2}$; $\frac{3}{2}$ & 0 &
$\Sigma_s \bar N$, $N \bar \Sigma_s$ & 2100 & 33 \\
$uud\,\, \bar u\bar u\bar s$, $udd\,\, \bar d\bar d\bar s$; & & &
$\Sigma^*_s \bar \Delta$, $\Delta \bar \Sigma^*_s$ & 2100 & 517 \\
$uus\,\, \bar u\bar d\bar d$, $dds\,\, \bar u\bar u\bar d$, &  & 1 &
$\Sigma_s \bar N$, $N \bar \Sigma_s$ & 2100 & 33 \\
$udd\,\, \bar u\bar u\bar s$, $uud\,\, \bar d\bar d\bar s$ &  & &
$\Sigma_s \bar \Delta$, $\Delta \bar \Sigma_s$ & 2100 & 325 \\
 &  & &
$\Sigma^*_s \bar N$, $N \bar \Sigma^*_s$ & 2100 & 225 \\
 &  & &
$\Sigma^*_s \bar \Delta$, $\Delta \bar \Sigma^*_s$ & 2100 & 517 \\
 & & 2 &
$\Sigma_s \bar \Delta$, $\Delta \bar \Sigma_s$ & 2108 & 317 \\
 & & &
$\Sigma^*_s \bar N$, $N \bar \Sigma^*_s$ & 2108 & 217 \\
 & & &
$\Sigma^*_s \bar \Delta$, $\Delta \bar \Sigma^*_s$ & 2108 & 509 \\
\hline
$uds\,\, \bar u\bar u\bar u$, $uds\,\, \bar d\bar d\bar d$, & $\frac{3}{2}$ & 0 &
$\Sigma^*_s \bar \Delta$, $\Delta \bar \Sigma^*_s$ & 2109 & 508 \\
$uuu\,\, \bar u\bar d\bar s$, $ddd\,\, \bar u\bar d\bar s$ & & 1, 2 &
$\Sigma_s \bar \Delta$, $\Delta \bar \Sigma_s$ & 2094 & 331 \\
 & & &
$\Sigma^*_s \bar \Delta$, $\Delta \bar \Sigma^*_s$ & 2094 &  523\\
 & & &
$\Lambda_s \bar \Delta$, $\Delta \bar \Lambda_s$ & 2094 & 254 \\
\hline
$uds\,\, \bar u\bar u\bar d$, $uds\,\, \bar u\bar d\bar d$, & $\frac{1}{2}$ & 0 &
$\Sigma_s \bar N$, $N \bar \Sigma_s$ & 2110 & 23 \\
$uud\,\, \bar u\bar d\bar s$, $udd\,\, \bar u\bar d\bar s$ & & &
$\Sigma^*_s \bar \Delta$, $\Delta \bar \Sigma^*_s$ & 2110 & 507 \\
 & & 1 &
$\Sigma_s \bar N$, $N \bar \Sigma_s$ & 2110 & 23\\
 & & &
$\Sigma_s \bar \Delta$, $\Delta \bar \Sigma_s$ & 2110 & 315 \\
 & & &
$\Sigma^*_s \bar N$, $N \bar \Sigma^*_s$, & 2110 & 215\\
& & &  $\Sigma^*_s \bar \Delta$, $\Delta \bar \Sigma^*_s$ & 2110 & 507 \\
& & & $\Lambda_s \bar \Delta$, $\Delta \bar \Lambda_s$ & 2110 & 238 \\
& & 2 & $\Sigma_s \bar \Delta$, $\Delta \bar \Sigma_s$ & 2126 & 299 \\
& & &
$\Sigma^*_s \bar N$, $N \bar \Sigma^*_s$ & 2126 & 199 \\
& & & $\Sigma^*_s \bar \Delta$, $\Delta \bar \Sigma^*_s$ & 2126 & 491 \\
& & &
$\Lambda_s \bar \Delta$, $\Delta \bar \Lambda_s$ & 2126 & 222 \\
\hline
\end{tabular}
\end{table}

\begin{table}
\caption{$qqQ\bar q\bar q\bar Q$, $q=u, d$, $Q=s$. Parameters of model: cutoff
$\Lambda=11.0$, $\Lambda_{qs, ss}=9.17$, gluon coupling constant $g=0.314$.
Quark masses $m_{u, d}=410\, MeV$, $m_s=557\, MeV$.}
\label{tab2}
\begin{tabular}{|c|c|c|c|c|c|}
\hline
Quark content & $I_3$ & $J$ & Baryonium & Mass (MeV) & Binding energy (MeV) \\
\hline
$uus\,\, \bar u\bar u\bar s$, $dds\,\, \bar d\bar d\bar s$; & 0; 2 & 0 &
$\Sigma_s \bar \Sigma_s$ & 2180 & 206 \\
$uus\,\, \bar d\bar d\bar s$, $dds\,\, \bar u\bar u\bar s$ & & &
$\Sigma^*_s \bar \Sigma^*_s$ & 2180 & 590 \\
 &  & 1 &
$\Sigma_s \bar \Sigma_s$ & 2179 &  207 \\
 &  & &
$\Sigma_s \bar \Sigma^*_s$, $\Sigma^*_s \bar \Sigma_s$ & 2179 &  399 \\
 &  & &
$\Sigma^*_s \bar \Sigma^*_s$ & 2179 &  591 \\
 & & 2 & $\Sigma_s \bar \Sigma^*_s$, $\Sigma^*_s \bar \Sigma_s$ & 2189 & 389 \\
 & & & $\Sigma^*_s \bar \Sigma^*_s$ & 2189 & 581 \\
\hline
$uus\,\, \bar u\bar d\bar s$, $dds\,\, \bar u\bar d\bar s$; & 1 & 0 &
$\Sigma_s \bar \Sigma_s$ & 2190 & 196 \\
$uds\,\, \bar u\bar u\bar s$, $uds\,\, \bar d\bar d\bar s$ & & &
$\Sigma^*_s \bar \Sigma^*_s$ & 2190 & 580 \\
 & & &
$\Sigma_s \bar \Lambda_s$, $\Lambda_s \bar \Sigma_s$ & 2190 & 119 \\
 & & 1 &
$\Sigma_s \bar \Sigma_s$ & 2190 & 196 \\
 & & &
$\Sigma_s \bar \Sigma^*_s$
$\Sigma^*_s \bar \Sigma_s$ & 2190 & 388 \\
 & & &
$\Sigma_s \bar \Lambda_s$, $\Lambda_s \bar \Sigma_s$ & 2190 & 119 \\
 & & &
$\Sigma^*_s \bar \Lambda_s$, $\Lambda_s \bar \Sigma^*_s$ & 2190 & 311 \\
 & & &
$\Sigma^*_s \bar \Sigma^*_s$ & 2190 & 580 \\
 & & 2 & $\Sigma_s \bar \Sigma^*_s$, $\Sigma^*_s \bar \Sigma_s$ & 2205 & 373 \\
 & & & $\Sigma^*_s \bar \Lambda_s$, $\Lambda_s \bar \Sigma^*_s$ & 2205 & 296 \\
 & & & $\Sigma^*_s \bar \Sigma^*_s$ & 2205 & 565 \\
\hline
$uds\,\, \bar u\bar d\bar s$ & 0 & 0 & $\Sigma_s \bar \Sigma_s$ & 2200 & 186 \\
 & & & $\Sigma_s \bar \Lambda_s$, $\Lambda_s \bar \Sigma_s$ & 2200 & 109 \\
 & & & $\Lambda_s \bar \Lambda_s$ & 2200 & 32 \\
 & & & $\Sigma^*_s \bar \Sigma^*_s$ & 2200 & 570 \\
 & & 1 & $\Sigma_s \bar \Sigma_s$ & 2200 & 186 \\
 & & & $\Sigma_s \bar \Lambda_s$, $\Lambda_s \bar \Sigma_s$ & 2200 & 109 \\
 & & & $\Sigma_s \bar \Sigma^*_s$, $\Sigma^*_s \bar \Sigma_s$ & 2200 & 378 \\
 & & & $\Sigma^*_s \bar \Lambda_s$, $\Lambda_s \bar \Sigma^*_s$ & 2200 & 301 \\
 & & & $\Lambda_s \bar \Lambda_s$ & 2200 & 32 \\
 & & & $\Sigma^*_s \bar \Sigma^*_s$ & 2200 & 570 \\
\hline
\end{tabular}
\end{table}

\begin{table}
\caption{The vertex functions and coefficients of Chew-Mandelstam functions.}
\label{tab3}
\begin{tabular}{|c|c|c|c|c|}
\hline
$i$ & $G_i^2(s_{kl})$ & $\alpha_i$ & $\beta_i$ & $\delta_i$ \\
\hline
& & & & \\
$0^+$ diquark & $\frac{4g}{3}-\frac{8gm_{kl}^2}{(3s_{kl})}$
& $\frac{1}{2}$ & $-\frac{1}{2}\frac{(m_k-m_l)^2}{(m_k+m_l)^2}$ & $0$ \\
& & & & \\
$1^+$ diquark & $\frac{2g}{3}$ & $\frac{1}{3}$
& $\frac{4m_k m_l}{3(m_k+m_l)^2}-\frac{1}{6}$
& $-\frac{1}{6}$ \\
& & & & \\
$0^-$ meson & $\frac{8g}{3}-\frac{16gm_{kl}^2}{(3s_{kl})}$
& $\frac{1}{2}$ & $-\frac{1}{2}\frac{(m_k-m_l)^2}{(m_k+m_l)^2}$ & $0$ \\
& & & & \\
$1^-$ meson & $\frac{4g}{3}$ & $\frac{1}{3}$
& $\frac{4m_k m_l}{3(m_k+m_l)^2}-\frac{1}{6}$
& $-\frac{1}{6}$ \\
& & & & \\
\hline
\end{tabular}
\end{table}

\end{document}